\documentclass{article}

\usepackage{amsmath}
\usepackage{amssymb}
\usepackage{graphicx,color}

\usepackage[T2A]{fontenc}
\usepackage[utf8]{inputenc}

\usepackage[russian,english]{babel}

\begin{document}

\title{Competing instabilities in wide-gap viscoelastic Taylor–Couette flow: Taylor vs. helical vortices}

\author{A. Proskurin\vspace{5mm}\\ \textit{Institute of Hydrodynamics SB RAS, Novosibirsk}\\ \textit{k210@list.ru} }

\date{}


\maketitle

\begin{abstract}
This paper presents a numerical study of the stability of a polymer solution flow between concentric cylinders with a rotating inner cylinder. The case of a small-radius inner cylinder is considered. The fluid motion is described using a specific case of the Kelvin–Voigt model, often referred to as the Oskolkov model. This model is applicable to very dilute polymer solutions, where the retardation time is much smaller than the characteristic time of the problem and elastic forces are much smaller than viscous forces. The stability of the steady-state motion is investigated using a fully nonlinear approach by means of direct numerical simulation of a perturbation introduced as finite-duration white noise. Depending on the Reynolds number, the perturbation either decays or grows. The critical Reynolds numbers obtained for both Newtonian and non-Newtonian fluids are found to be in agreement with the predictions of the linear theory. It is also shown that an increase in the elastic forces makes helical perturbations more dangerous than their axisymmetric counterparts.
\end{abstract}

\section{Introduction}

Polymer solutions are frequently encountered in industrial and domestic applications. Compared to Newtonian fluids, polymer molecules have an elongated shape, which imparts elastic properties to the fluids containing them. These elastic properties significantly influence the onset of instabilities in fluid flows.

One of the important model problems for studying fluid properties is the Couette--Taylor flow. This problem is of great importance as it allows the investigation of fluid properties using a relatively simple experiment. The flow regimes and instabilities of Couette--Taylor flow for Newtonian fluids are well understood (see, for example, \cite{chossat2012couette} and \cite{yaglom2012hydrodynamic}), which facilitates the identification of various phenomena's effects on the flow.

In the work \cite{proskurin2025stability}, the stability of Couette--Taylor flow of a viscoelastic fluid was investigated within the case of the zeroth-order Kelvin--Voigt model. The outer cylinder was fixed, and only the inner cylinder rotated. It was found that for a small inner cylinder radius, the helical mode can be more dangerous compared to the axisymmetric Taylor mode. This behavior differs radically from the Newtonian fluid case and could be easily detected experimentally. To confirm this finding, the present work provides the results of direct numerical simulation of perturbation development in a viscoelastic fluid.

The problem formulation and governing equations are presented in Section \ref{a37_formulation}, the numerical method and computational setup are described in Section \ref{a37_methods}, the results are presented in Section \ref{a37_results}, and discussion and conclusions are given in Section \ref{a37_conclusion}.

\section{Problem formulaion}\label{a37_formulation}

\begin{figure}[t]
\begin{center}
\includegraphics{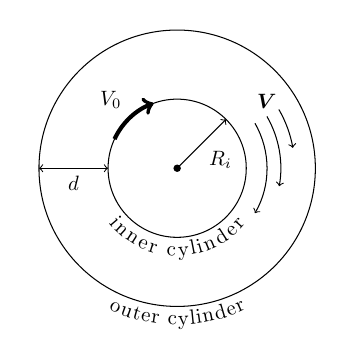}
\end{center}
\caption{A cross-section of the coaxial cylinders normal to their axis}\label{a37_TC-Geometry}
\end{figure}

Consider the flow between two infinite coaxial cylinders, shown schematically in cross-section in Fig. \ref{a37_TC-Geometry}. The outer cylinder is fixed, while the inner cylinder (of radius $R_i$) rotates with a constant angular velocity; the velocity of its surface is denoted by $V_0$ and is used as the velocity scale. The gap width $d$ is used as the length scale. The dimensionless inner cylinder radius $\xi = \frac{R_i}{d}$ completely defines the geometry of the problem. Based on the results of linear stability theory from \cite{proskurin2025stability}, the cylinder length was set to $21.3$, and the parameter $\xi = 0.049$. For these parameters, the helical perturbation is more dangerous than the axisymmetric Taylor mode.

The motion of an incompressible fluid is described by the equations:
\begin{equation}
\label{a37_NSEquation_1}
\begin{aligned}
\frac{\partial \boldsymbol{v}}{\partial t} + \left(\boldsymbol{v}\cdot\nabla \right)&\boldsymbol{v} = -\nabla p + \nabla \cdot \boldsymbol{\sigma},\\
\nabla \cdot &\boldsymbol{v} = 0,
\end{aligned}
\end{equation}
where $\boldsymbol{v}$ is the fluid velocity, $p$ is the pressure, and $\boldsymbol{\sigma}$ is the stress tensor. The stress tensor can be expressed in terms of the rate-of-strain tensor $\boldsymbol{D} = \frac{1}{2}(\nabla \boldsymbol{v} + \nabla \boldsymbol{v}^T)$ using a constitutive relation derived from specific assumptions about the fluid properties. For a viscoelastic fluid, one of the simplest models is the zeroth-order Kelvin--Voigt model proposed in \cite{oskolkov1988initial}, which takes the form:
\begin{equation}
\label{a37_const_eq}
\boldsymbol{\sigma} = 2(\nu \boldsymbol{D}+\kappa \frac{\partial D}{\partial t}).
\end{equation}
Here, $\nu$ is the fluid viscosity, $\kappa$ is the viscoelastic coefficient ($\kappa \geq 0$), representing the retardation time. In this case, the equations becomes:
\begin{equation}
\label{a37_NSEquation_2}
\begin{aligned}
\frac{\partial \boldsymbol{v}}{\partial t} + \left(\boldsymbol{v}\cdot\nabla \right)&\boldsymbol{v} = -\nabla p +\nu \Delta \boldsymbol{v} +  \kappa \frac{\partial}{\partial t} \Delta \boldsymbol{v},\\
\nabla \cdot &\boldsymbol{v} = 0,
\end{aligned}
\end{equation}
where the characteristic parameters are the Reynolds number $Re = \frac{V_0 d}{\nu}$ and the dimensionless retardation time $K = \frac{\kappa}{d^2}$. The standard no-slip and impermeability conditions are imposed on the cylinders, as in the Newtonian case:
\begin{equation}
\label{a37_bound_cond}
\boldsymbol{v} = 0.
\end{equation}

\section{Numerical method}\label{a37_methods}
The calculations were performed using the Nektar++ framework \cite{moxey2020nektar++}, which implements the numerical scheme proposed in \cite{karniadakis1991high}. The solver for incompressible viscous flow was extended to compute the additional terms required by the Oskolkov model. Let us consider a sequence of time instants $t_0, t_1, \ldots, t_{n-1}, t_n, t_{n+1}$. Uusing a first-order finite difference scheme, an intermediate velocity $\tilde{\boldsymbol{v}}$ is found from:
\begin{equation}
\label{article37.VelCorr_1st_inter_vel}
\frac{\tilde{\boldsymbol{v}}-\boldsymbol{v}_n}{\delta t} = -\left(\boldsymbol{v}_n \nabla \right)\boldsymbol{v}_n.
\end{equation}
where $ (\boldsymbol{v}_n \cdot \nabla) \boldsymbol{v}_n$ is the advection term. The second intermediate velocity $\hat{\boldsymbol{v}}$ is then defined by:
\begin{equation}
\label{article37.VelCorr_2nd_inter_vel}
\frac{\hat{\tilde{\boldsymbol{v}}}-\tilde{\boldsymbol{v}}}{\delta t} = - \nabla p_{n+1}.
\end{equation} 
Using the continuity equation $\nabla \cdot \hat{\boldsymbol{v}} = 0$, the Poisson equation for pressure is obtained:
\begin{equation}
\label{article37.VelCorr_Poissin_eq}
\Delta p_{n+1} = \nabla \left(\frac{\tilde{\boldsymbol{v}}}{\delta t}\right)
\end{equation}
with appropriate boundary conditions. In the final step of the algorithm, the following Helmholtz equation is solved:
\begin{equation}
\label{article37.VelCorr_n+1_vel}
\left(\nu+\frac{\kappa}{\delta t}\right)\Delta \boldsymbol{v}_{n+1}-\frac{\boldsymbol{v}_{n+1}}{\delta t} = \frac{\kappa}{\delta t}\Delta \boldsymbol{v}_n - \frac{\tilde{\boldsymbol{v}}}{\delta t}  + \nabla p_{n+1},
\end{equation}
which yields the velocity $\boldsymbol{v}_{n+1}$ at the new time step.

\begin{figure}[t]
\begin{center}
\includegraphics[width=0.6\textwidth]{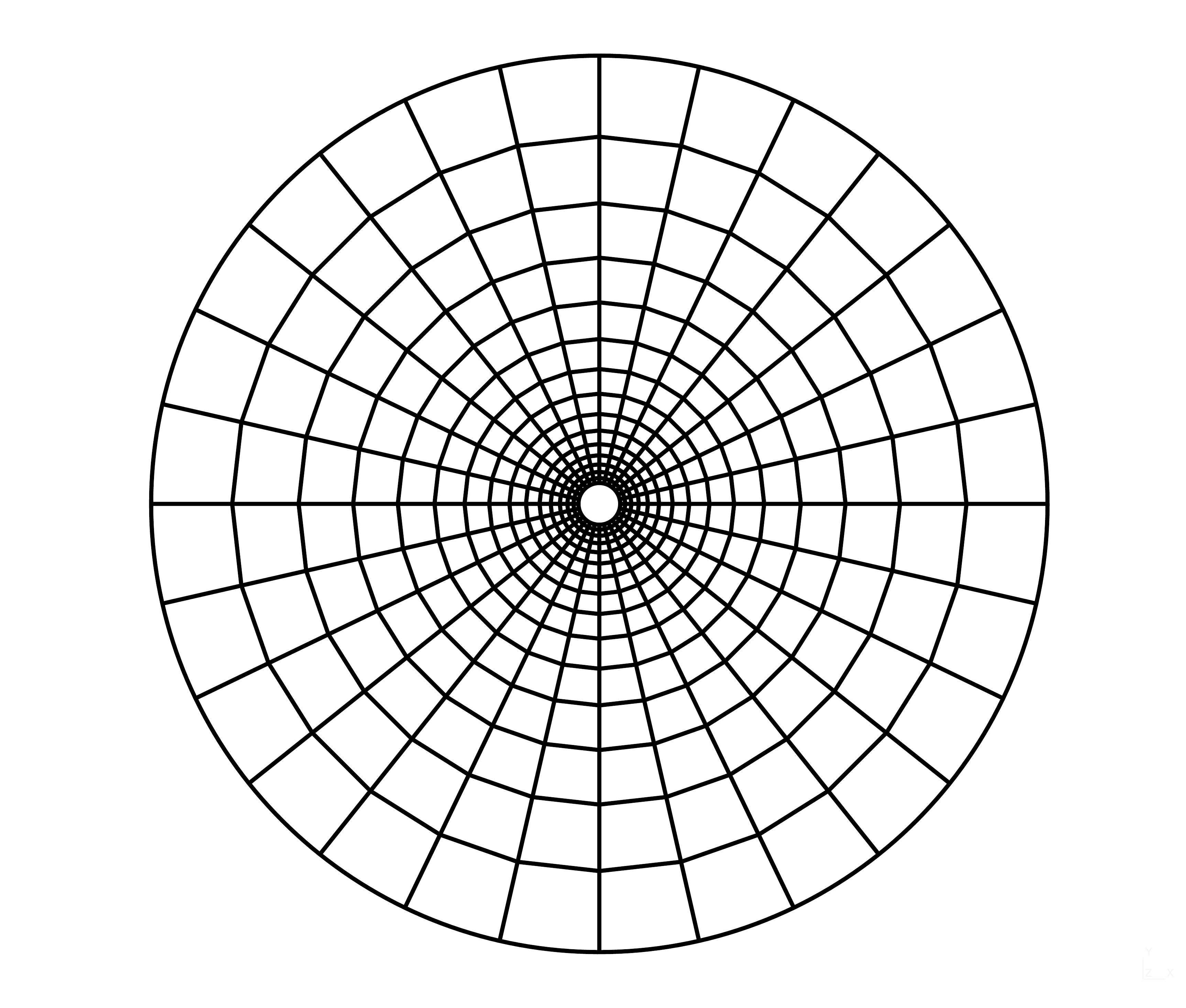}
\end{center}
\caption{The mesh}\label{a37_Mesh7}
\end{figure}

\begin{figure}[t]
\begin{center}
\includegraphics[width=0.6\textwidth]{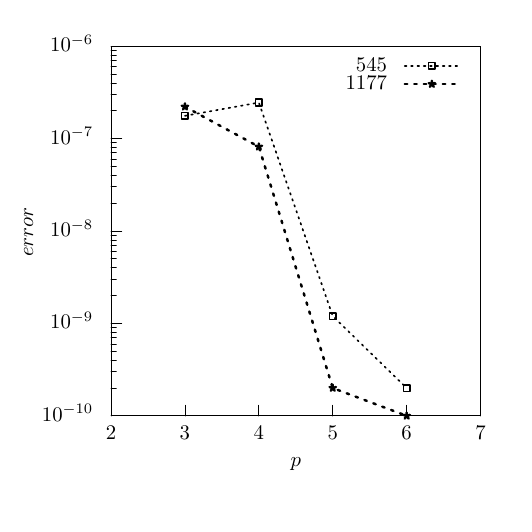}
\end{center}
\caption{Convergence depending on polynomial order $p$}\label{a37_convergence}
\end{figure}

The three-dimensional velocity field was computed using a $2.5D$ approach: a Fourier spectral representation was employed along the $z$-direction, while the cross-sectional flow field was discretised using the mesh shown in Figure~\ref{a37_Mesh7}. This mesh contained $545$ elements. A finer mesh with $1177$ elements was also used for comparison. To assess convergence, a series of calculations were performed at $Re=250$ with non-axisymmetric initial conditions, varying the polynomial order $p$, the number of Fourier modes $nz$, and the time step $\Delta t$. Figure \ref{a37_convergence} shows the dependence of the error on the polynomial order $p$, where the error was computed as the difference between the velocity at a fixed point for a given $p$ and the velocity at the maximum $p$. Based on these results, the following parameters were selected for the main calculations using the $545$-element mesh: $p=5$, $nz=50$, $\Delta t=0.025$, ensuring convergence to $10^{-6} \cdot V_0$.

\section{Results of computations}\label{a37_results}

\begin{figure}[t]
\begin{center}
\includegraphics[width=0.6\textwidth]{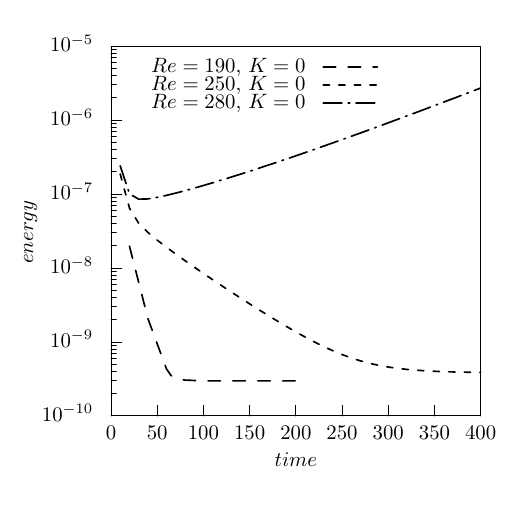}
\end{center}
\caption{Perturbation energy vs. time for $K=0$}\label{a37_energy-K0p0}
\end{figure}

\begin{figure}[t]
\begin{center}
\includegraphics[width=0.6\textwidth]{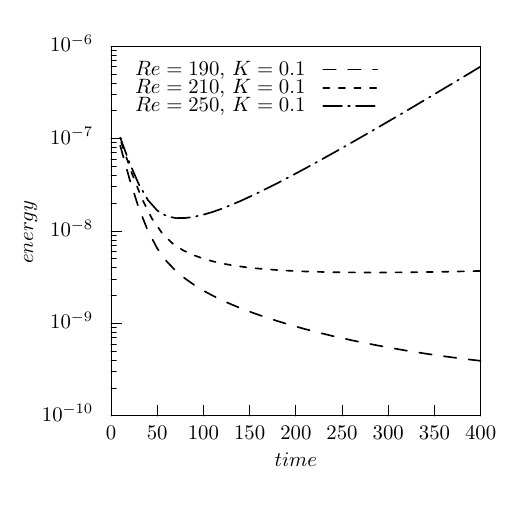}
\end{center}
\caption{Perturbation energy vs. time for $K=0.1$}\label{a37_energy-K0p1}
\end{figure}

\begin{figure}[t]
\begin{center}
\includegraphics[width=0.6\textwidth]{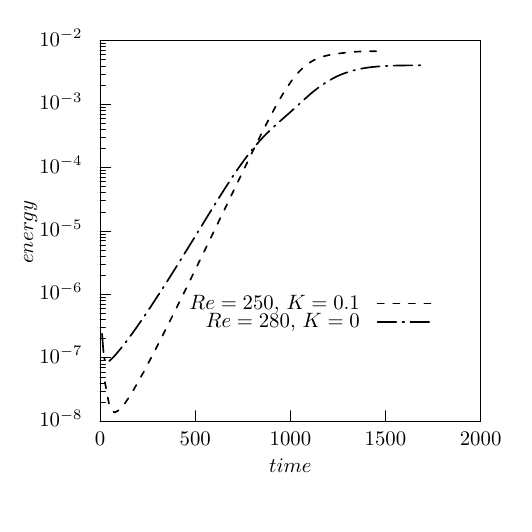}
\end{center}
\caption{Long-time perturbation energy evolution for $K=0$, $Re=280$ and $K=0.1$, $Re=250$}\label{a37_energy-long-time}
\end{figure}

\begin{figure}[t]
\begin{center}
\begin{tabular}{cc}
\includegraphics[width=0.45\textwidth]{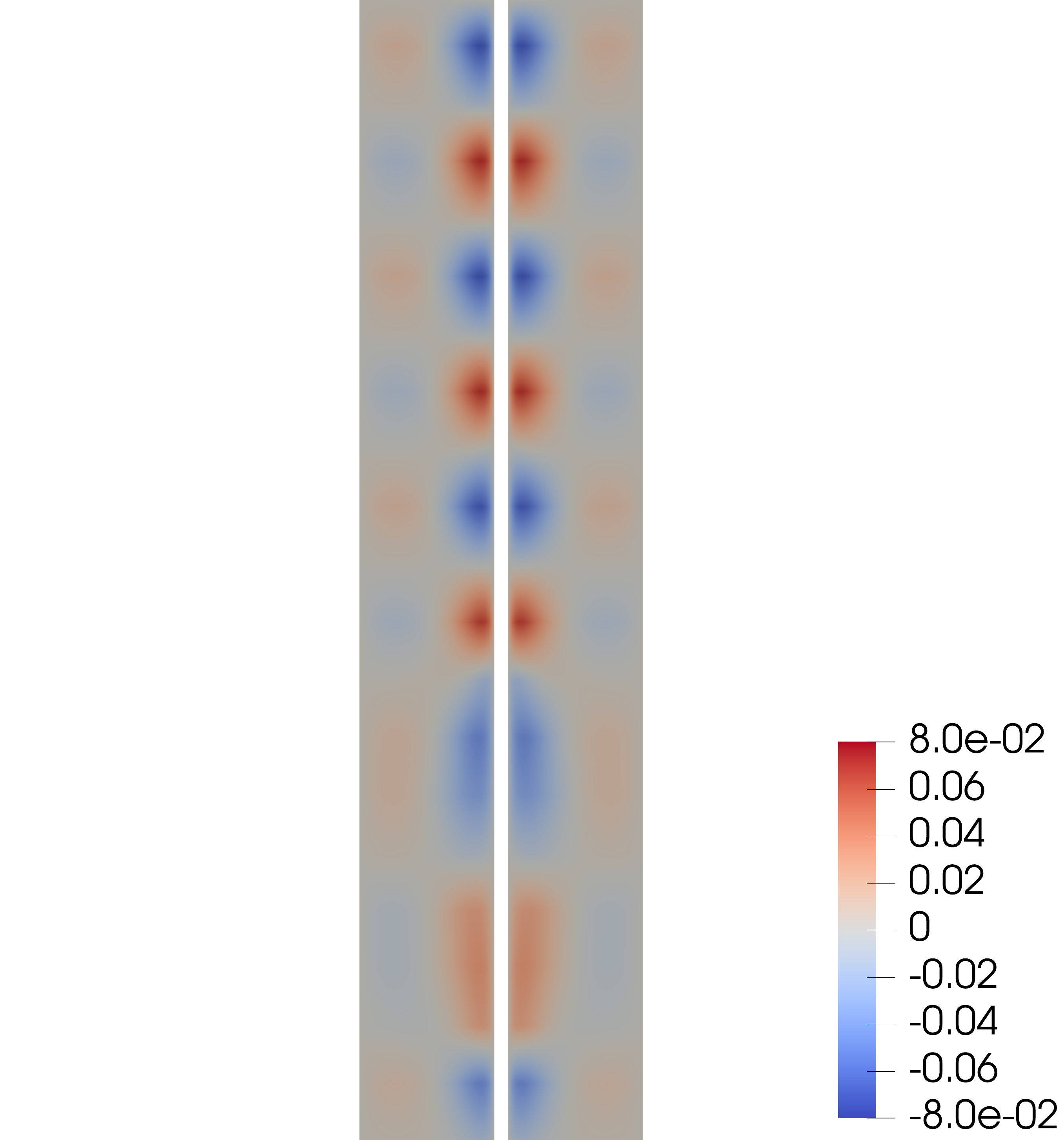} & \includegraphics[width=0.45\textwidth]{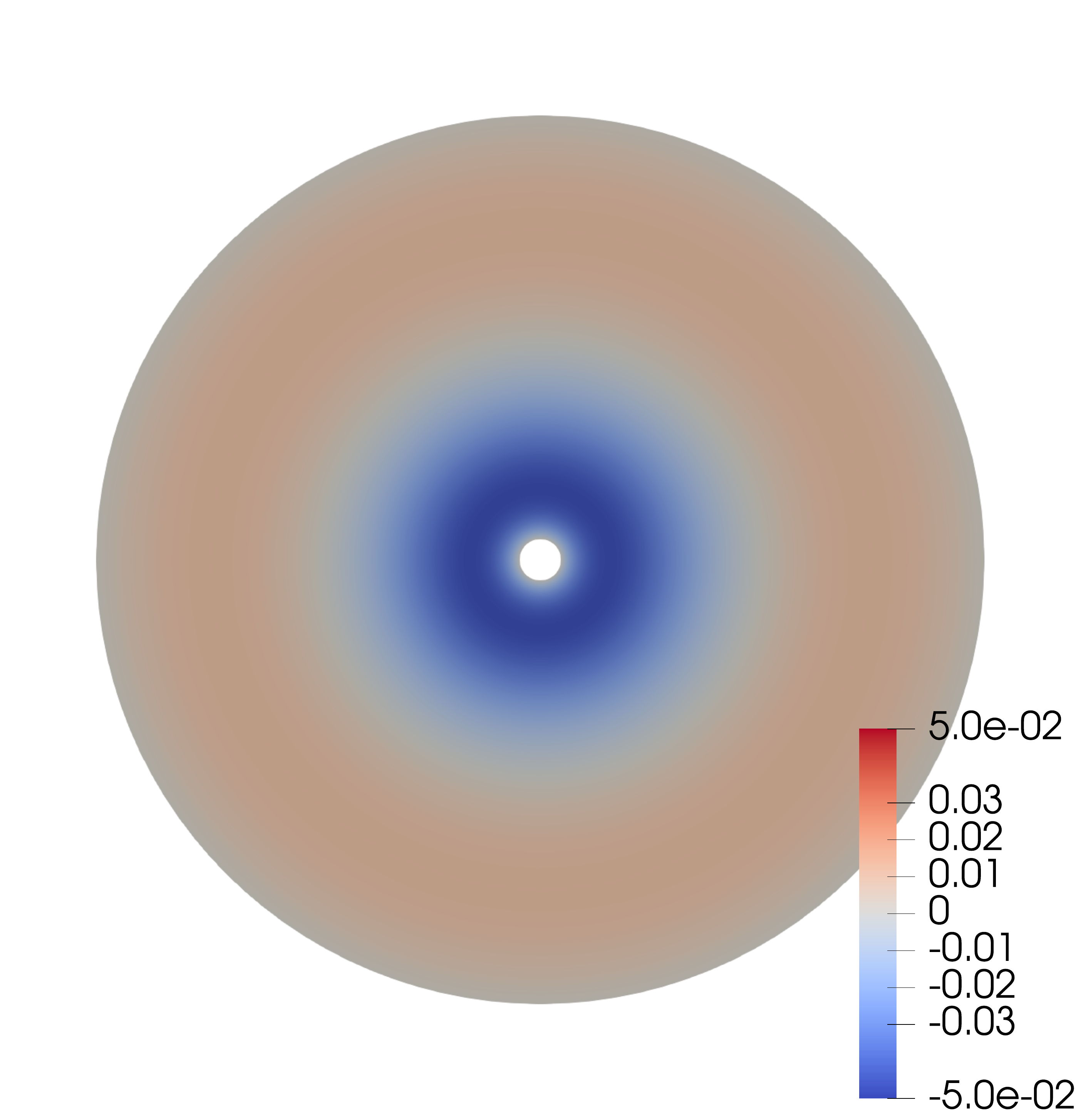}  \\
(a) & (b)  \\
\includegraphics[width=0.45\textwidth]{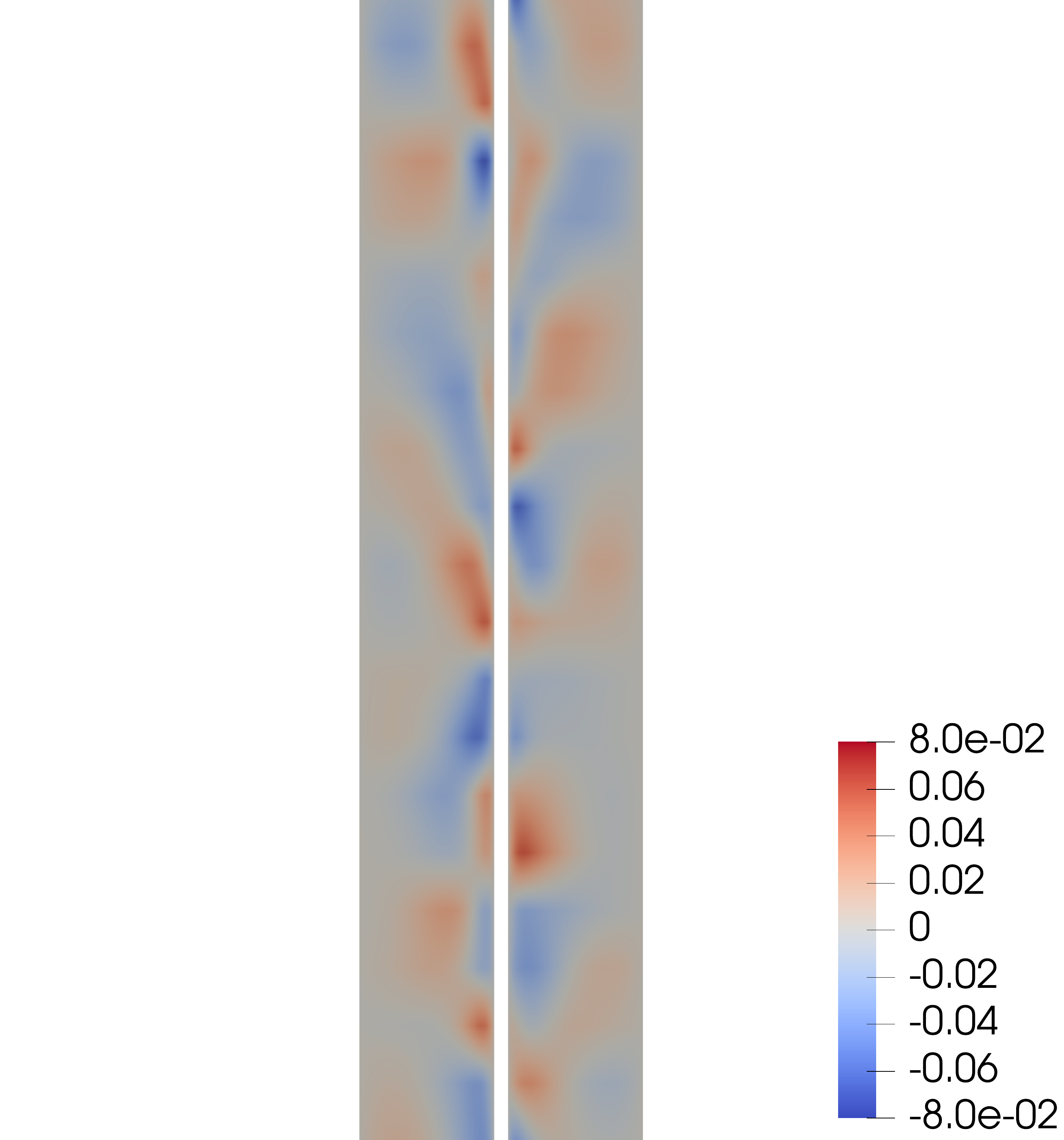} & \includegraphics[width=0.45\textwidth]{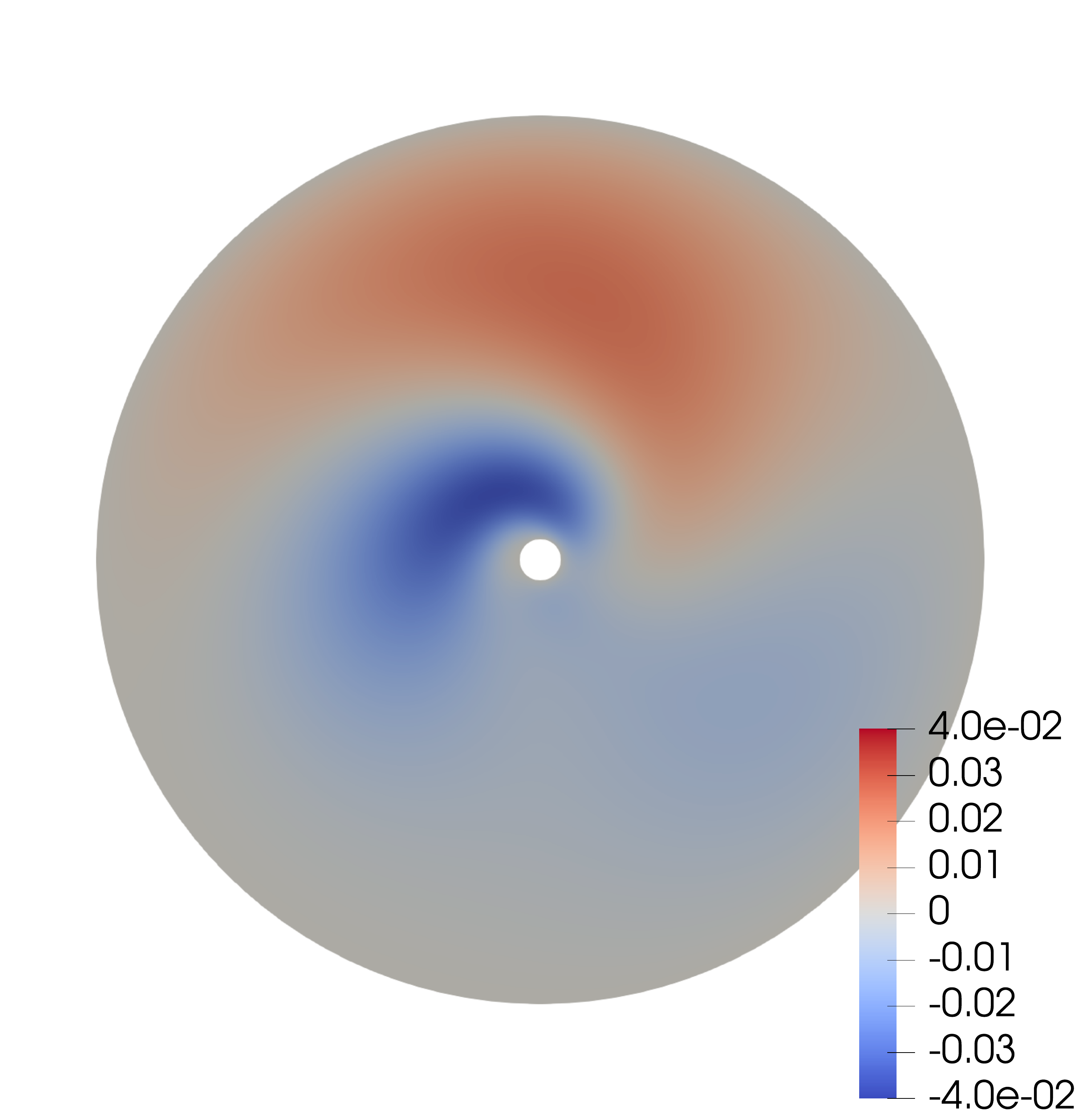}  \\
(c) & (d)
\end{tabular}

\end{center}
\caption{Axial velocity of the axisymmetric perturbation at $Re=280$ and $K=0$ in the longitudinal (a) and transverse (b) sections; axial velocity of the helical perturbation at $Re=250$ and $K=0.1$ in the longitudinal (c) and transverse (d) sections}\label{a37_re250patterns}
\end{figure}

During the calculations, the time evolution of perturbations superimposed on the steady-state flow was studied. The integration of the governing equations was performed in time. The steady-state flow was used as the initial condition, and the perturbation was imposed as white noise using the standard Nektar++ feature. The noise amplitude and the time interval over which it was imposed were varied without affecting the critical Reynolds numbers. A series of calculations was performed for selected parameter sets at $K=0$ and $K=0.1$ for various Reynolds numbers. The results are presented in Figures~\ref{a37_energy-K0p0} and~\ref{a37_energy-K0p1} as the time dependence of the $L_2$ norm of the perturbation energy. For $K=0$, at $Re=190$, the perturbation energy drops sharply to a level presumably determined by numerical errors. At $Re=250$, the energy also decreases, while at $Re=280$, it increases. These observations are consistent with linear theory, which predicts a critical Reynolds number of $Re_{*}=265$ (computed using the methods described in \cite{proskurin2025stability}). For $K=0.1$, shown in Figure~\ref{a37_energy-K0p1}, three energy evolution curves are presented: at $Re=190$ the perturbation decays, at $Re=210$ it settles at a constant level, and at $Re=250$ it grows. These results correspond to the linear theory critical Reynolds number $Re_{*}=202$. For the cases $K=0, Re=280$ and $K=0.1, Re=250$, the long-time evolution of the perturbations was examined over a time interval sufficient for the energy to reach a constant level, which should correspond to Taylor vortices or the helical vortices, respectively. The corresponding energy curves are shown in Figure~\ref{a37_energy-long-time}, while Figure~\ref{a37_re250patterns} shows the axial velocity component fields in the longitudinal and transverse cross-sections.

\section{Discussion and conclusions}\label{a37_conclusion}

The paper presents the results of a direct numerical simulation of the laminar-turbulent transition in the Couette--Taylor flow of a zeroth-order Kelvin--Voigt viscoelastic fluid. The calculations were performed for a fixed value of the inner cylinder radius $\xi=0.049$, which is small compared to the outer one. Two values of the dimensionless retardation time were investigated: $K=0$, corresponding to the case of a viscous fluid, and $K=0.1$, for which the viscoelastic effects in the context of stability are sufficiently large, as shown in \cite{proskurin2025stability}. The evolution of a perturbation generated by finite-amplitude random noise was considered. At Reynolds numbers below the critical value predicted by linear theory, the perturbations decayed, while above it, they grew. Thus, linear theory can be used to predict the loss of stability in Couette--Taylor flow in the case of a wide gap between the cylinders, for both ordinary viscous fluids and polymer solutions. An increase in the influence of elasticity, as found in \cite{proskurin2025stability}, leads to a change in the secondary flow regime that emerges instead of the steady Couette--Taylor flow. Instead of Taylor vortices, helical vortices appear, which for a Newtonian fluid are observed only when the outer cylinder is counter-rotating \cite{chossat2012couette}. In other words, the forces of molecular elasticity give rise to a new instability mode that competes with the Taylor mode.

\bibliographystyle{unsrt}
\bibliography{reference37}

\end{document}